\documentclass[%
 reprint,
 superscriptaddress,
 amsmath,amssymb,
 aps,
 prl,
]{revtex4-1}

\usepackage{graphicx}% Include figure files
\usepackage{dcolumn}% Align table columns on decimal point
\usepackage{bm}% bold math
\usepackage{colortbl}
\usepackage[table]{xcolor}
\usepackage{makecell}
\usepackage[mathscr]{euscript}
\usepackage[table]{xcolor}
\usepackage{multirow}
\usepackage{braket}
\usepackage[pagewise]{lineno}
\usepackage{hyperref}
\usepackage{lipsum}

\begin{document}

%\preprint{APS/123-QED}

\title{Engineering ferroelectricity in monoclinic hafnia}

\author{Hong Jian Zhao}
 \affiliation{Key Laboratory of Material Simulation Methods and Software of Ministry of Education, College of Physics, Jilin University, Changchun 130012, China}
 \affiliation{Key Laboratory of Physics and Technology for Advanced Batteries (Ministry of Education), College of Physics, Jilin University, Changchun 130012, China}
 \affiliation{International Center of Future Science, Jilin University, Changchun 130012, China}
\author{Yuhao Fu}
 \affiliation{Key Laboratory of Material Simulation Methods and Software of Ministry of Education, College of Physics, Jilin University, Changchun 130012, China}
 \author{Longju Yu}
 \affiliation{Key Laboratory of Material Simulation Methods and Software of Ministry of Education, College of Physics, Jilin University, Changchun 130012, China}
 \author{Yanchao Wang}
 \email{wyc@calypso.cn}
 \affiliation{Key Laboratory of Material Simulation Methods and Software of Ministry of Education, College of Physics, Jilin University, Changchun 130012, China}
 \affiliation{State Key Laboratory of Superhard Materials, College of Physics, Jilin University, Changchun 130012, China}
\author{Yurong Yang}
\affiliation{National Laboratory of Solid State Microstructures and Jiangsu Key Laboratory of Artificial Functional Materials, Department of Materials Science and Engineering, Nanjing University, Nanjing 210093, China}
\author{Laurent Bellaiche}
\affiliation{Physics Department and Institute for Nanoscience and Engineering, University of Arkansas, Fayetteville, Arkansas 72701, USA}
\author{Yanming Ma}
\email{mym@jlu.edu.cn}
 \affiliation{Key Laboratory of Material Simulation Methods and Software of Ministry of Education, College of Physics, Jilin University, Changchun 130012, China}
 \affiliation{International Center of Future Science, Jilin University, Changchun 130012, China}
 \affiliation{State Key Laboratory of Superhard Materials, College of Physics, Jilin University, Changchun 130012, China}

\date{\today}

\begin{abstract}
Ferroelectricity in the complementary metal–oxide semiconductor (CMOS)-compatible hafnia (HfO$_2$) is crucial for the fabrication of high-integration nonvolatile memory devices. However, the capture of ferroelectricity in HfO$_2$ requires the stabilization of thermodynamically-metastable orthorhombic or rhombohedral phases, which entails the introduction of defects (\textit{e.g.}, dopants and vacancies) and pays the price of crystal imperfections, causing unpleasant wake-up and fatigue effects. Here, we report a theoretical strategy on the realization of robust ferroelectricity in HfO$_2$-based ferroelectrics by designing a series of epitaxial (HfO$_2$)$_1$/(CeO$_2$)$_1$ superlattices. The advantages of the designated ferroelectric superlattices are defects free, and most importantly, on the base of the thermodynamically stable monoclinic phase of HfO$_2$. Consequently, this allows the creation of superior ferroelectric properties with an electric polarization $>$25 $\mu$C/cm$^2$ and an ultralow polarization-switching energy barrier at $\sim$2.5 meV/atom. Our work may open an entirely new route towards the fabrication of high-performance HfO$_2$ based ferroelectric devices. 
\end{abstract}

\maketitle

\noindent
\textit{Introduction. --} Ferroelectric materials being compatible with complementary metal–oxide semiconductor (CMOS) are crucial for the fabrication of high-integration nonvolatile memory devices~\cite{phasehfo2zro2,hfo2lesson}. The hafnia (HfO$_2$) -- extensively used in the CMOS devices~\cite{phasehfo2zro2} -- became one such material since the discovery of ferroelectricity in Si-doped HfO$_2$~\cite{hfo2first}. However, the ferroelectric phases of HfO$_2$ are thermodynamically metastable over a broad temperature range ({\it e.g.}, from 0~$^\circ$C to 1500~$^\circ$C) and at ambient pressure  -- the stable phase being the  monoclinic $P2_1/c$ phase~\cite{hfo2phase2}. The strategies for stabilizing the ferroelectric phases mostly lie in the introduction of defects ({\it e.g.}, dopants and vacancies), under special fabrication conditions ({\it e.g.}, suitable film thickness and epitaxial strain)~\cite{phasehfo2zro2,qi2020,liu2023,ri2021,mono2021,hfo2surface,hfo2surface2,shiliu2019hfo2,liuprb,hfo2stable}. 
Such strategies yield the fabrication of various HfO$_2$-based ferroelectrics, represented by Y-doped HfO$_2$~\cite{xu2021,yun2022,hfo2ori,shimizu2015}, La-doped HfO$_2$~\cite{lahfo2,lahfo22,lahfo23}, Hf$_{1-x}$Ce$_x$O$_2$~\cite{hfceo2,hfceo22,hfceo23,hfceo24,hfceo25}, Hf$_{0.5}$Zr$_{0.5}$O$_2$~\cite{hfzro2,hfzro22,hfzro23,wei2018,nukala2020,nukala2021r3m,hfo2he}, HfO$_2$/ZrO$_2$ superlattices~\cite{mixfeafe,hfo2zro2,hfo2zro22} and Hf(Zr)$_{1+x}$O$_2$~\cite{hfo2barrier2} with $Pca2_1$ or $R3m$ space group.
Despite that HfO$_2$-based ferroelectrics were successfully achieved, these materials often suffer from the imperfections such as crystal defects and mixed non-ferroelectric phases (see \textit{e.g.}, Ref.~\cite{phasehfo2zro2}). Such imperfections are detrimental to the ferroelectric cycling stabilities in HfO$_2$-based ferroelectrics, causing wake-up and fatigue effects~\cite{phasehfo2zro2,mono2021,cycl1,cycl2}.

Here, we explore the possibility for engineering ferroelectricity in defect-free monoclinic HfO$_2$ by symmetry analysis and first-principles simulations. We show that such ferroelectricity can be achieved by creating the (HfO$_2$)$_1$/(CeO$_2$)$_1$ superlattices, where Hf and Ce ions are appropriately ordered. Unlike the previously-reported HfO$_2$-based ferroelectrics, our proposed ferroelectric (HfO$_2$)$_1$/(CeO$_2$)$_1$ superlattices are on the base of the thermodynamically-stable $P2_1/c$ HfO$_2$; engineering ferroelectricity in these superlattices is thus natural, allowing defect-irrelevant HfO$_2$-based ferroelectrics. Promisingly, our designated (HfO$_2$)$_1$/(CeO$_2$)$_1$ superlattice presents superior ferroelectric properties with a sizable electric polarization ($>25$~$\mu$C/cm$^2$) and an ultralow polarization-switching energy barrier at $\sim$2.5 meV/atom. \\

\begin{figure*}[t!]
\centering
\includegraphics[width=1\linewidth]{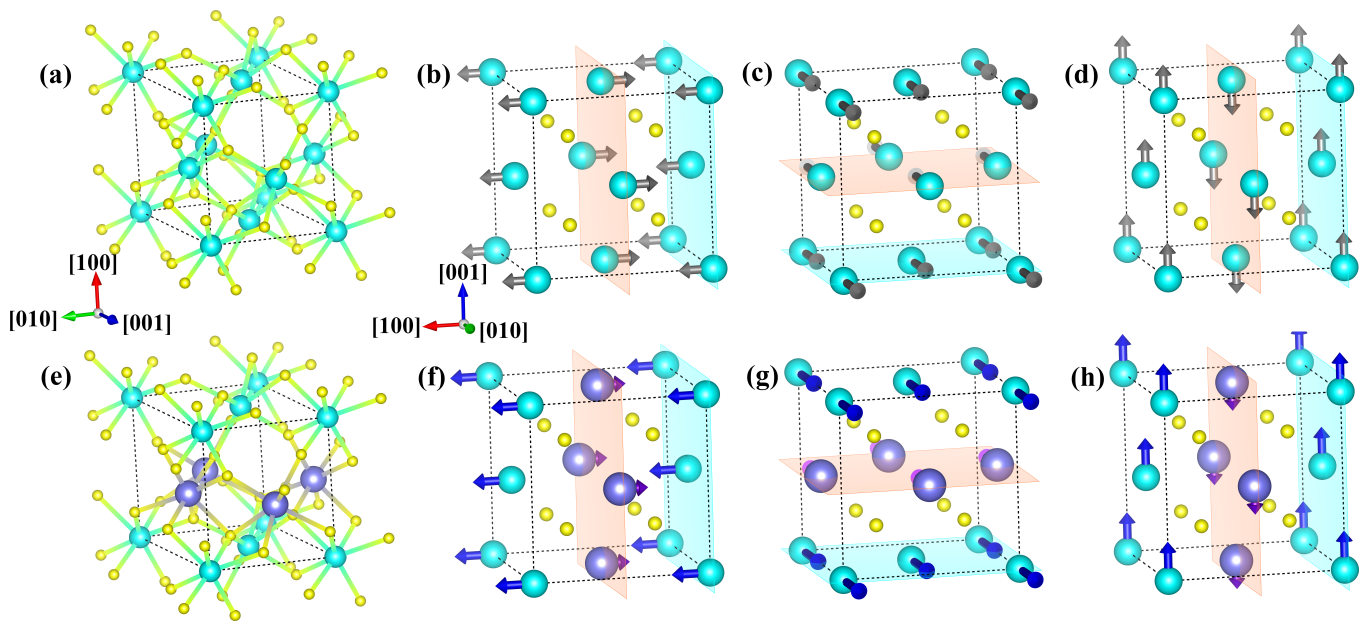}
\caption{\label{fig:hfo2hif} Panel (a): Schematization of the $P2_1/c$ phase of HfO$_2$. Panels (b)--(d): Three anti-polar motions of Hf ions in the $P2_1/c$ phase of HfO$_2$. Panel (e): The [100]-oriented (HfO$_2$)$_1$/($X$O$_2$)$_1$ superlattice obtained by replacing a sequence of Hf ions (of the $P2_1/c$ HfO$_2$ oxide) by $X$ ions; such replacements are ordered with respect to the (100) crystallographic plane. In reality, the positions of Hf, $X$ and O in the (HfO$_2$)$_1$/($X$O$_2$)$_1$ superlattice will be adjusted, compared with those of Hf and O in $P2_1/c$ HfO$_2$. In panel (e), we merely sketch the formation of the [100]-oriented (HfO$_2$)$_1$/($X$O$_2$)$_1$ superlattice (from $P2_1/c$ HfO$_2$), without demonstrating the adjustments of the ionic positions. Panels (f) and (h): The atomic displacements in the [100]-oriented (HfO$_2$)$_1$/($X$O$_2$)$_1$ superlattice. Panel (g): The atomic displacements in the [001]-oriented superlattice. The Hf, $X$ and O ions are represented by cyan, purple and yellow spheres, respectively. The displacements of Hf or $X$ ions are denoted by grey, blue or purple arrows -- the displacements of O being not shown. In panels (a) and (e), the [100] and [001] axes form a non-90$^\circ$ monoclinic angle; in panels (b)--(d) and (f)--(h), the [100], [010] and [001] axes are perpendicular to each other.}
\end{figure*}

\noindent
\textit{Creating ferroelectricity in the monoclinic phase. --} The lattice parameters $(a,b,c)$ of the bulk $P2_1/c$ HfO$_2$ -- obtained by first-principles calculations  -- are (5.08, 5.16, 5.25)~\AA, where $a$, $b$ and $c$ are along the [100], [010] and [100] crystallographic directions, respectively [see Fig.~\ref{fig:hfo2hif}(a)]. The [100] and [001] directions exhibit a monoclinic angle of $\sim99.6^\circ$. Symmetry analysis indicates that the Hf sublattice in this $P2_1/c$ phase presents three types of anti-polar motions [see Figs.~\ref{fig:hfo2hif}(b)--(d)], compared with the high-symmetric $Fm\bar{3}m$ phase~\cite{hfo2longju}. In these motions, we can identify two types of crystallographic planes, colored  cyan and orange in Figs.~\ref{fig:hfo2hif}(b)--(d). The Hf ions within the cyan or orange plane are displaced along the same direction, while those between the cyan and orange planes are moved oppositely. 
In Fig.~\ref{fig:hfo2hif}(b), we depict one anti-polar motion involving the displacements of Hf ions long $[100]$ and $[\bar{1}00]$ directions. The Hf ions displaced along $[100]$ and those displaced along $[\bar{1}00]$ are stacked along the crystallographic [100] direction; more vividly, the cyan and orange planes in Fig.~\ref{fig:hfo2hif}(b) are alternately aligned along the crystallographic [100] orientation. In the same stacking mode, the displacements of Hf ions can also occur along $[001]$ and $[00\bar{1}]$ orientations [see Fig.~\ref{fig:hfo2hif}(d)]. Besides, the $P2_1/c$ phase exhibits another anti-polar motion [see Fig.~\ref{fig:hfo2hif}(c)], where the displacements of Hf ions are along $[010]$ and $[0\bar{1}0]$ orientations and the stacking (of the $[010]$-displaced and $[0\bar{1}0]$-displaced Hf ions) is occurring along the crystallographic [001] direction. In these three cases, the overall displacements of Hf ions along opposite directions are identical in magnitude, compensating with each other ({\it i.e.}, no electric polarization).

To engineer ferroelectricity in $P2_1/c$ phase of HfO$_2$, a possible strategy is to create the (HfO$_2$)$_1$/($X$O$_2$)$_1$ superlattice with $X$ being different from Hf. In this superlattice, the Hf and $X$ ions should be stacked in such a way that Hf and $X$ ions are displaced along $\pm\alpha$ and $\mp\alpha$ directions ($\alpha$ being $[100]$, $[010]$ or $[001]$)~\footnote{The $-[uvw]$ direction should be understood as the $[\bar{u}\bar{v}\bar{w}]$ crystallographic direction.}, respectively [see Figs.~\ref{fig:hfo2hif}(f)--\ref{fig:hfo2hif}(h)]. Essentially, the (HfO$_2$)$_1$/($X$O$_2$)$_1$ superlattices can be obtained via the replacements of some specific Hf ions in $P2_1/c$ HfO$_2$ by $X$ ions, as exemplified by Fig.~\ref{fig:hfo2hif}(e). Under these circumstances, the Hf and $X$ ions are displaced, albeit oppositely, with non-compensated magnitudes, yielding net off-center displacements and an electric polarization.

As shown in Figs.~\ref{fig:hfo2hif}(f) and~\ref{fig:hfo2hif}(h), the [100]-oriented (HfO$_2$)$_1$/($X$O$_2$)$_1$ superlattice~\footnote{In the following, the [$uvw$]-oriented (HfO$_2$)$_1$/($X$O$_2$)$_1$ superlattice is referred to as the superlattice where the Hf and $X$ ions are ordered by layer along the crystallographic [$uvw$] direction.} allows electric polarization along $\pm[100]$ and $\pm[001]$ directions. Figure~\ref{fig:hfo2hif}(g) sketches the displacements of Hf and $X$ ions in the [001]-oriented (HfO$_2$)$_1$/($X$O$_2$)$_1$ superlattice. This superlattice gains an electric polarization along $\pm[010]$ direction. 
As for the [010]-oriented superlattice, no net off-center displacements can be expected. Our aforementioned arguments are confirmed by the symmetry analysis: the [100]- and [001]-oriented (HfO$_2$)$_1$/($X$O$_2$)$_1$ superlattices exhibit $Pc$ and $P2_1$ space groups, respectively, both of which are compatible with ferroelectricity. The [010]-oriented (HfO$_2$)$_1$/($X$O$_2$)$_1$ superlattice, on the other hand, are centrosymmetric and non-ferroelectric. Our strategy for engineering ferroelectricity in (HfO$_2$)$_1$/($X$O$_2$)$_1$ superlattice can basically be linked with the notion of hybrid improper ferroelectricity (HIF) that was previously developed to describe the ferroelectricity in perovskites superlattices and Ruddlesden-Popper compounds~\cite{hif1,hif2,hif3,hif4}.\\

\begin{figure}[t!]
\centering
\includegraphics[width=1.0\linewidth]{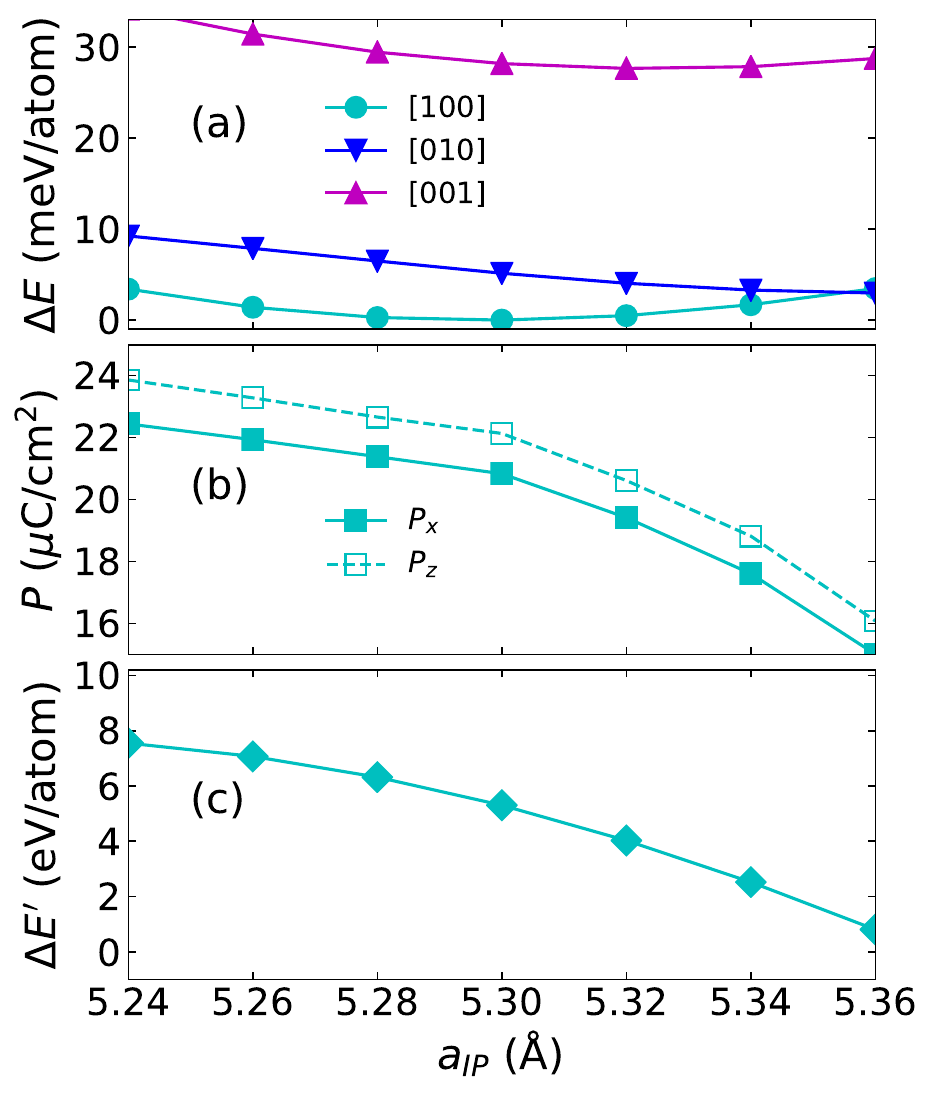}
\caption{\label{fig:hfo2ceo2} Various physical quantities of the (HfO$_2$)$_1$/(CeO$_2$)$_1$ superlattice as a function of the in-plane lattice parameter $a_{IP}$. Panel (a): The relative energy $\Delta E$ of the [100]-, [010]- and [001]-oriented superlattices. Panel (b): The electric polarization $P$ of the [100]-oriented superlattices. Panel (c): The polarization-switching energy barrier $\Delta E^\prime$ for the [100]-oriented superlattices.}
\end{figure}

\noindent
\textit{Energetics of the (HfO$_2$)$_1$/(CeO$_2$)$_1$ superlattices. --} Our aforementioned discussion suggests that stabilizing the [100]- or [001]-oriented (HfO$_2$)$_1$/($X$O$_2$)$_1$ superlattice is the key to engineering ferroelectricity in it. In view of this, we decide to find appropriate $X$O$_2$ oxide so that it can combine with $P2_1/c$ HfO$_2$ and form the [100]- or [001]-oriented superlattice. We recall that the $P2_1/c$ HfO$_2$ can be seen as the derivative of the fluorite-type $Fm\bar{3}m$ HfO$_2$~\cite{phasehfo2zro2}. Searching the Materials Project database~\cite{materialsproject,materialsproject2} and following Ref.~\cite{hfo2lesson}, we identify CeO$_2$ as the candidate for $X$O$_2$. 
The CeO$_2$ oxide is fluorite-structured with the $Fm\bar{3}m$ space group~\footnote{The first-principles calculated lattice parameter is 5.40~\AA~for bulk CeO$_2$} over a wide temperature spectrum (e.g., from 0$^\circ$C to 2500~$^\circ$C)~\cite{ceo2,ceo2zro2n1,ceo2zro2n2}. Earlier work shows that CeO$_2$ can form solid solutions with HfO$_2$, yielding HfO$_2$-based ferroelectrics (see e.g., Refs.~\cite{hfceo2,hfceo22,hfceo23,hfceo24,hfceo25}). The shared prototype structure (i.e., fluorite type) of $P2_1/c$ HfO$_2$ and $Fm\bar{3}m$ CeO$_2$ implies the possibility for achieving the (HfO$_2$)$_1$/(CeO$_2$)$_1$ superlattice. Experimentally, this kind of short-period superlattices can be epitaxially grown on appropriate substrates. To accommodate the [100]- or [001]-oriented (HfO$_2$)$_1$/(CeO$_2$)$_1$ superlattice, the cubic $<$100$>$-oriented substrate with appropriate in-plane lattice parameter $a_{IP}$ should be selected for such epitaxial growth.

Now, we mimic the (HfO$_2$)$_1$/(CeO$_2$)$_1$ superlattice grown on various substrates ({\it i.e.}, with various $a_{IP}$) by first-principles calculations. We examine the [100]-, [010]- and [001]-oriented (HfO$_2$)$_1$/(CeO$_2$)$_1$ superlattices, since superlattices of these types can geometrically match the cubic $<$100$>$-oriented substrate. For each of the superlattice, we fix its in-plane lattice vectors to $(a_{IP}, 0, 0)$ and $(0, a_{IP}, 0)$, and relax its out-of-plane lattice vector and atomic positions. Figure~\ref{fig:hfo2ceo2}(a) shows the energetics of the [100]-, [010]- and [001]-oriented (HfO$_2$)$_1$/(CeO$_2$)$_1$ superlattices as a function of $a_{IP}$. The [100]-oriented (HfO$_2$)$_1$/(CeO$_2$)$_1$ superlattice is more stable than the [010]- and [001]-oriented cases over a broad $a_{IP}$ range ({\it e.g.}, from 5.24~\AA~to 5.34~\AA). These in-plane lattice parameters correspond to (i) compressive strains ranging from $-3.0\%$ to $-1.1\%$ with respect to 5.40~\AA~of the bulk $Fm\bar{3}m$ CeO$_2$, and (ii) tensile strains ranging from $0.8\%$ to $2.7\%$ with respect to $(b+c)/2=5.20$~\AA~of the bulk $P2_1/c$ HfO$_2$. In particular, the [100]-oriented (HfO$_2$)$_1$/(CeO$_2$)$_1$ superlattice with $a_{IP}=5.30$~\AA~roughly corresponds to the $-1.9\%$ compressively-strained CeO$_2$ and the $1.9\%$ tensilely-strained HfO$_2$. Compared with their bulk phases, the HfO$_2$ and CeO$_2$ in the [100]-oriented (HfO$_2$)$_1$/(CeO$_2$)$_1$ superlattices are moderately strained. Such moderate strain conditions imply that our proposed [100]-oriented (HfO$_2$)$_1$/(CeO$_2$)$_1$ superlattice can likely be grown by epitaxy.\\

\begin{figure*}[t!]
\centering
\includegraphics[width=1\linewidth]{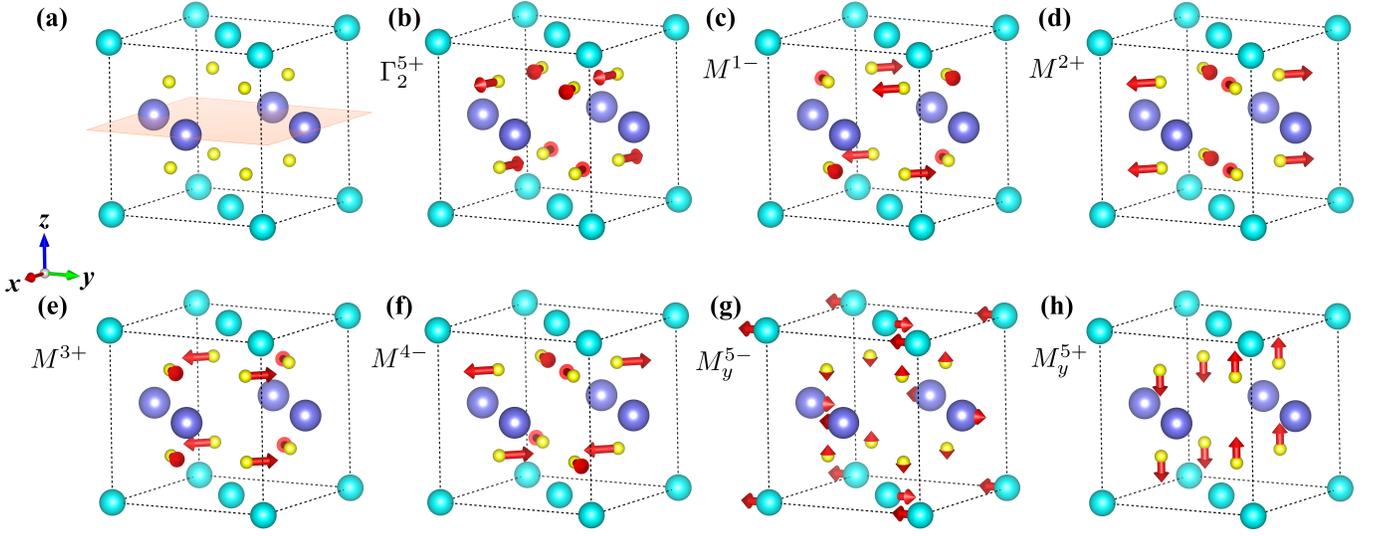}
\caption{\label{fig:hif100} The schematizations of the non-polar atomic motions in the [100]-oriented (HfO$_2$)$_1$/(CeO$_2$)$_1$ superlattice ($Pc$ space group). Panel (a): The $P4/mmm$ reference phase of (HfO$_2$)$_1$/(CeO$_2$)$_1$. Panels (b)--(h): The non-polar atomic motions. The Hf, Ce and O ions are denoted by cyan, purple and yellow spheres. The Cartesian directions of our coordinate system are labelled by ``$x$'', ``$y$'' and ``$z$''. The crystallographic [100] direction in $P2_1/c$ HfO$_2$ roughly corresponds to the $z$ axis. The atomic displacements are shown by red arrows. In the $M^{5-}_y$ motion, the displacements contributed by Ce and O ions are negligible compared with those from Hf ions; for displaying clarity, we enlarge the magnitudes of Ce's and O's displacements in panel (g).}
\end{figure*}

\noindent
\textit{Landau theory for HIF in the (HfO$_2$)$_1$/($Ce$O$_2$)$_1$ superlattices. --} Prior to studying the ferroelectric behaviors of the [100]-oriented (HfO$_2$)$_1$/(CeO$_2$)$_1$, we develop the Landau theory for describing the ferroelectricity in the (HfO$_2$)$_1$/(CeO$_2$)$_1$ superlattices. To this end, we start from the cubic $Fm\bar{3}m$ HfO$_2$ and create a tetragonal (HfO$_2$)$_1$/(CeO$_2$)$_1$ superlattice ($P4/mmm$ space group), where the Hf and Ce ions are ordered by layer along $z$ direction [see Fig.~\ref{fig:hif100}(a)]. With respect to the tetragonal phase, we identify a sequence of non-polar atomic motions [see Fig.~S1 of the Supplementary Material (SM)] that may be hosted by the [100]- and [001]-oriented (HfO$_2$)$_1$/(CeO$_2$)$_1$ superlattices. Apart from these non-polar motions, there are three polar motions in (HfO$_2$)$_1$/(CeO$_2$)$_1$ superlattices, namely, $P_\chi$  ($\chi=x,y,z$) being associated with the electric polarization along $\chi$ direction.
These atomic motions exhibit a variety of trilinear couplings shown in Eq.~(S1) of the SM. As for the [100]-oriented (HfO$_2$)$_1$/(CeO$_2$)$_1$ superlattice, symmetry analysis suggests the following atomic motions: $O_{100}\equiv(P_x,P_z,\Gamma^{5+}_2,M^{1-},M^{2+},M^{3+},M^{4-},M^{5-}_y,M^{5+}_y)$, with the non-polar motions schematized in Fig.~\ref{fig:hif100}. The trilinear couplings regarding the [100]-oriented superlattice are $H_{100} =  \alpha_1 M^{2+}M^{4-}P_z + \alpha_2 M^{3+}M^{1-}P_z + \alpha_3  M^{5-}_y M^{5+}_y P_z + \kappa  P_x P_z \Gamma^{5+}_2 + \beta_2  P_x M^{5-}_y M^{2+} + \beta_3 P_x M^{5-}_y  M^{3+}   + \beta_6 P_x M^{5+}_y M^{4-}   + \beta_7 P_x M^{5+}_y M^{1-} +  \lambda_3 \Gamma^{5+}_2 M^{5-}_y M^{4-}   - \lambda_4 \Gamma^{5+}_2 M^{5-}_y M^{1-} +   \lambda_7 \Gamma^{5+}_2 M^{5+}_y M^{2+} - \lambda_8  \Gamma^{5+}_2 M^{5+}_y M^{3+}$. Of particular interest are given by $H^\mathrm{Polar}_{100}=\alpha_1 M^{2+}M^{4-}P_z + \alpha_2 M^{3+}M^{1-}P_z + \alpha_3  M^{5-}_y M^{5+}_y P_z  + \beta_2  P_x M^{5-}_y M^{2+} + \beta_3 P_x M^{5-}_y  M^{3+}   + \beta_6 P_x M^{5+}_y M^{4-}   + \beta_7 P_x M^{5+}_y M^{1-}$. These terms are $P_\chi X_1 X_2$-type couplings ($X_1$ and $X_2$ being two non-polar motions) and imply that the combination of $X_1$ and $X_2$ non-polar motions leads to electric polarization $P_\chi$. For instance, the coexistence of $M^{2+}$ and $M^{4-}$ non-polar motions [see Figs.~\ref{fig:hif100}(d) and~\ref{fig:hif100}(f)] indicates the emergence of electric polarization along $z$ direction, as suggested by the $\alpha_1 M^{2+}M^{4-}P_z$ term.

We move on to discuss the polarization switching path in the [100]-oriented (HfO$_2$)$_1$/(CeO$_2$)$_1$ superlattice. To this end, we take our aforementioned $O_{100}$ state as our initial state, and identify the possible final states when switching the polarization from $(P_x,P_z)$ to $(-P_x,-P_z)$. The $H_{100}$ discussed above indicates that reversing the polarization will change some of the non-polar motions and maintain others. This can be demonstrated as follows, taking $\alpha_2 M^{3+}M^{1-}P_z$ and $\beta_7 P_x M^{5+}_y M^{1-}$ terms as examples. The $\alpha_2 M^{3+}M^{1-}P_z$ term implies that reversing $P_z$ will flip either $M^{1-}$ or $M^{3+}$ (but not both). Working with $\beta_7 P_x M^{5+}_y M^{1-}$, this causes two possible consequences: (i) if $M^{3+}$ is not flipped ($M^{1-}$ being flipped), $M^{5+}_y$ will not be flipped either, and (ii) if $M^{3+}$ is flipped ($M^{1-}$ being not flipped), $M^{5+}_y$ will be flipped as well. Furthermore, the $\kappa  P_x P_z \Gamma^{5+}_2$ term implies that $\Gamma^{5+}_2$ will be unchanged when switching $(P_x,P_z)$ to $(-P_x,-P_z)$. Following this logic, we identify two possible final states $O^{\prime}_{100}$ and $O^{\prime\prime}_{100}$ with respect to $O_{100}$. Apart from the reversed $P_x$ and $P_z$, the $O^{\prime}_{100}$ state showcases reversed $M^{2+}$, $M^{3+}$ and $M^{5+}_y$, while the $O^{\prime\prime}_{100}$ state exhibits reversed $M^{1-}$, $M^{4-}$ and $M^{5-}_y$. The $O^{\prime}_{100}$ and $O^{\prime\prime}_{100}$ final states together with the $O_{100}$ initial state suggest the $O_{100}$-$O^\prime_{100}$ and $O_{100}$-$O^{\prime\prime}_{100}$ polarization switching paths for the [100]-oriented (HfO$_2$)$_1$/(CeO$_2$)$_1$ superlattice~\footnote{The analysis regarding the [001]-oriented (HfO$_2$)$_1$/(CeO$_2$)$_1$ superlattice can be found in Section II of the SM.}. \\

\noindent
\textit{Ferroelectricity in the [100]-oriented (HfO$_2$)$_1$/(CeO$_2$)$_1$ superlattices. --}
We move on to determine the polarization switching behavior for the [100]-oriented (HfO$_2$)$_1$/(CeO$_2$)$_1$ superlattice with $a_{IP}=5.30$~\AA. Regarding this superlattice, the $O_{100}$-$O^{\prime}_{100}$ (respectively, $O_{100}$-$O^{\prime\prime}_{100}$) switching path indicates the intermediate $P2/c$ (respectively, $P2_1/c$) phase for polarization switching~\footnote{The intermediate $P2/c$ (respectively, $P2_1/c$) phase can be obtained by the linear interpolation between the $O_{100}$ initial state and the $O^{\prime}_{100}$ (respectively, $O^{\prime\prime}_{100}$) final state. The ratio for such a linear interpolation is 1:1.}. The energy barriers for the polarization switching via $O_{100}$-$O^{\prime}_{100}$ and $O_{100}$-$O^{\prime\prime}_{100}$ paths are $\sim$13.6 and $\sim$77.2 meV/atom, respectively, obtained by first-principles self-consistent calculations (without structural relaxations). The nudged elastic band (NEB) algorithm~\footnote{See \url{https://theory.cm.utexas.edu/vtsttools/neb.html} for the nudged elastic band algorithm. This algorithm determines the possible transition paths between initial and final states, with the minimum energy barriers.} further decreases the energy barrier (regarding the $O_{100}$-$O^{\prime}_{100}$ path) to $\sim$5.3 meV/atom. We also examine the effect of $a_{IP}$ on the polarization-switching barrier in [100]-oriented (HfO$_2$)$_1$/(CeO$_2$)$_1$ superlattices [see Fig.~\ref{fig:hfo2ceo2}(b)]. Varying $a_{IP}$ from 5.24 to 5.34~\AA~reduces the barriers from $\sim$7.5 to $\sim$2.5 meV/atom. The energy barriers for polarization switching in [100]-oriented (HfO$_2$)$_1$/(CeO$_2$)$_1$ are far lower than $\sim$40 meV/atom in the $Pca2_1$ phase of HfO$_2$~\cite{hfo2barrier}. In particular, the ultralow energy barrier of $\sim$2.5 meV/atom in (HfO$_2$)$_1$/(CeO$_2$)$_1$ superlattice ($a_{IP}$ being 5.34~\AA) is also significantly reduced, compared with $\sim$7.6 meV/atom in the rhombohedral Hf$_{1.08}$O$_2$~\cite{hfo2barrier2}. 
This implies that the [100]-oriented epitaxial (HfO$_2$)$_1$/(CeO$_2$)$_1$ superlattices enable the polarization switching via ultralow coercive electric field.

We next compute the electric polarization for the [100]-oriented (HfO$_2$)$_1$/(CeO$_2$)$_1$ superlattices. By symmetry, the [100]-oriented (HfO$_2$)$_1$/(CeO$_2$)$_1$ enables the $P_x$ and $P_z$ components of the polarization. Figure~\ref{fig:hfo2ceo2}(c) shows the electric polarization of the [100]-oriented (HfO$_2$)$_1$/(CeO$_2$)$_1$ as a function of $a_{IP}$. At $a_{IP}=5.30$~\AA, the $P_x$ and $P_z$ are $\sim$20.8 and $\sim$22.1 $\mu$C/cm$^2$, yielding a total electric polarization of $\sim$30.4 $\mu$C/cm$^2$. Varying $a_{IP}$ from 5.24 to 5.34~\AA~decreases the total electric polarization from $\sim$32.8 to $\sim$25.8 $\mu$C/cm$^2$. The electric polarization values in the [100]-oriented (HfO$_2$)$_1$/(CeO$_2$)$_1$ superlattices are thus sizable, being comparable to those observed in various HfO$_2$-based ferroelectrics [{\it e.g.}, $\sim$22 $\mu$C/cm$^2$ in Hf(Zr)$_{1+x}$O$_2$~\cite{hfo2barrier2}, $\sim$34 $\mu$C/cm$^2$ in Hf$_{0.5}$Zr$_{0.5}$O$_2$~\cite{wei2018}, and $\sim$50 $\mu$C/cm$^2$ in Y-doped HfO$_2$~\cite{yun2022}]. \\

\noindent
\textit{Summary. --}  In summary, we demonstrate by symmetry analysis that the Hf sublattice in the monoclinic phase of HfO$_2$ presents three types of anti-polar motions. These anti-polar motions enable the creation of electric polarization in this monoclinic phase by forming the [100]- or [001]-oriented (HfO$_2$)$_1$/($X$O$_2$)$_1$ superlattice ($X\neq\mathrm{Hf}$). Our first-principles calculations further predict that the [100]-oriented (HfO$_2$)$_1$/(CeO$_2$)$_1$ superlattice may be experimentally accessible via the epitaxial growth of it on cubic $<$100$>$-oriented substrates. The epitaxial (HfO$_2$)$_1$/(CeO$_2$)$_1$ superlattice showcases ferroelectricity that is tunable by varying its in-plane lattice parameters ($a_{IP}$). To be specific, varying $a_{IP}$ from 5.24 to 5.34~\AA~results in (i) the decrease of polarization from $\sim$32.8 to $\sim$25.8 $\mu$C/cm$^2$, and (ii) the reduction of polarization-switching barrier from $\sim7.5$ to $\sim2.5$ meV/atom. The designated ferroelectric superlattice ($a_{IP}=5.34$~\AA) showcases a sizable electric polarization $>25$~$\mu$C/cm$^2$ and an ultralow energy barrier at $\sim2.5$ meV/atom, promising for the design of nonvolatile memory devices ({\it e.g.}, ferroelectric random-access memory) with high integration and low power cost. \\

\noindent
\textit{Acknowledgements. --}  We acknowledge the support from the National Key Research and Development Program of China (Grant No. 2022YFA1402502) and the National Natural Science Foundation of China (Grants No. T2225013, No. 12274174, No. 52288102, and No. 12034009). L. B. acknowledges support from the Vannevar Bush Faculty Fellowship (VBFF) from the Department of Defense and Award No. DMR-1906383 from the National Science Foundation Q-AMASE-i Program (MonArk NSF Quantum Foundry). 
The authors thank the support from the high-performance computing center of Jilin University. We also thank Prof. Zuhuang Chen at Harbin Institute of Technology (Shenzhen) and Prof. Yingfen Wei at Fudan University for valuable discussion.

%\bibliography{ref}

\end{document}